\documentstyle[12pt]{article}
\begin{document}

\begin{flushright}
TAUP-2250-95\\
hep-th/9508057
\end{flushright}

\baselineskip 24pt
\newcommand{\be}{\begin{equation}}
\newcommand{\ee}{\end{equation}}
\newcommand{\leqx}{\,\raisebox{-1.0ex}{$\stackrel{\textstyle <}
{\sim}$}\,}
\def\geqx{\,\raisebox{-1.0ex}{$\stackrel{\textstyle >}{\sim}$}\,}

\begin{center}
{\bf Information Loss in Quantum Gravity Without Black Holes}
\\N.Itzhaki \footnote{Email Address:sanny@halo.tau.ac.il}
\\Raymond and Beverly Sackler Faculty of  Exact Sciences
\\School
 of Physics and Astronomy
\\Tel Aviv University, Ramat Aviv, 69978, Israel
\end{center}
\begin{abstract}
We use the  weak field
approximation to show that information is lost in principle in quantum
gravity.
\end{abstract}
\newpage
{\bf 1.Introduction}

In fields theories which do not involve gravitation, the information that a
macroscopic system has is the $log$ of the number of orthogonal states with
the same macroscopic properties (E, V, N, ...).
There is, however, a hidden assumption in this definition.
The hidden assumption is that one can distinguish at least in principle
 between all the microscopic orthogonal states which define the
same macroscopic properties.
Fortunately, in field theories which do not involve gravitation the
assumption is correct.

In this paper we examine this  assumption in the context of quantum gravity.
We suggest that the assumption is incorrect and that there is information
loss in quantum gravity.
This result, in itself, is not new; Hawking has shown in \cite{ha}
that information is lost in the context of black holes.
Since then there have been many suggestions how to recover the information
lost in a
black hole, most of them  based on the argument that one of the
assumptions that Hawking made on quantum gravity is incorrect.
However, it seems fair to say that the description of quantum black holes is
still a puzzle.
Further more, the meaning of the black hole entropy, $\frac{A}{4}$,
is not clear and
the connection between black hole entropy and entropy in fields theories
which do not involve gravitation is not well understood \cite{be}.
Our new point is that we do not consider black holes, but use only general
properties of quantum gravity and the weak field approximation; in
particular there is no horizon in our system.
This is quite surprising since it is usually claimed that the basic reason
for information loss in black holes is the horizon.
The general properties of quantum gravity which we assume are the
following:

1-At large distances, general relativity equation is a good approximation
to quantum gravity.
We denote the minimal scale for which general relativity equation is a good
approximation to quantum gravity by $x_{c}$.
In other words, at distances much larger then $x_{c}$ the first order of the
gravitational effects can be described by general relativity.

2-At  large distances, quantum gravity can be described by means of local
quantum field theory.
In other words, if one explores the theory only at large scales one finds
only local effects.
Let us denote the minimal scale for which the theory is local by $x_{d}$.
In particular, any measurement at scales larger then $x_{d}$ is carried out by
a local
interaction  between the measurement apparatus
and the measured system.

Notice that these two assumptions are the easiest way to describe the
correspondence principle, which we know to exist, between quantum gravity
and general relativity and between quantum gravity and local quantum field
theory.
In fact this correspondence principle is the only experimental data that
we have on quantum gravity.
Still, those are only assumptions on quantum gravity and not general properties
of quantum gravity, since there are more complicated ways to
describe the     correspondence principle.
For example, one can speculate that $x_{d}$ is not just a constant of nature,
but a function of the state of the system.
Notice further that those assumptions are usually made in any attempt to
construct a quantum theory of gravitation, where usually $x_{c}\approx x_{d}
\approx
L_{p}$ ( where $ L_{p}$ is Planck lenght); in particular, Hawking made those
assumptions in his description of black hole evaporation \cite{ha1}.
In our discussion we do not  assume that $x_{c}$ and/or $x_{d}$ are of the
order of  $ L_{p}$ but we only assume that they are finite.
The outline of the paper is as follows: In sec.2 we present the system to be
discussed and we show that the information assumption is correct
in the absence of gravitation.
In sec.3 we show that
 in the presence of gravitation, one can not distinguish between all the
orthogonal states of the system, therefore, the information assumption is
incorrect.

{\bf 2.The system}

Let us describe the system that we want to investigate.
There are $n^3$ objects in the system, each object is constructed from
$m^3$ cells whose size  is $a$,
\be a\gg max[x_{c},x_{d}].\ee
The cells are attached to each other so the size of each object is $ma$,
(Figure~1).
\begin{figure}
\begin{picture}(300,210)(0,0)
\put(50,20){\line(1,0){160}}
\put(210,180){\line(0,-1){160}}
\put(50,180){\line(0,-1){80}}
\put(70,180){\line(0,-1){80}}
\put(90,180){\line(0,-1){80}}
\put(110,180){\line(0,-1){80}}
\put(130,180){\line(0,-1){80}}
\put(50,180){\line(1,0){80}}
\put(50,160){\line(1,0){80}}
\put(50,140){\line(1,0){80}}
\put(50,120){\line(1,0){80}}
\put(50,100){\line(1,0){80}}
\put(150,180){.}
\put(170,180){.}
\put(190,180){.}
\put(50,40){.}
\put(50,60){.}
\put(50,80){.}
\put(220,90){\vector(0,-1){70}}
\put(220,110){\vector(0,1){70}}
\put(60,190){\vector(1,0){10}}
\put(60,190){\vector(-1,0){10}}
\put(55,195){a}
\put(225,95){ma}
\end{picture}
\caption{The object}
\end{figure}
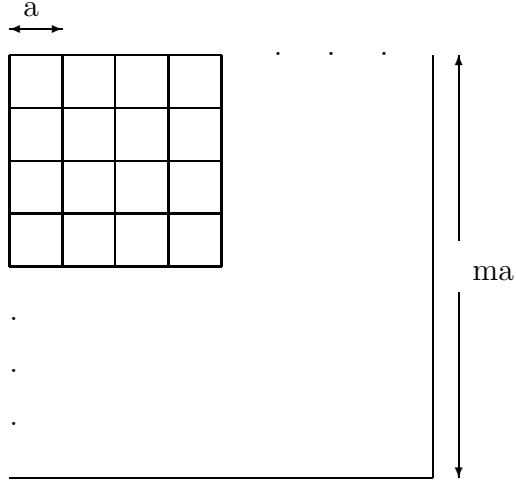
There are also $n^3$ huge cells whose size  is $b$ ($b\gg
ma$); the huge cells are also attached to each other, so the size of the system
is
$nb$, (Figure~2).
At each little cell there is exactly one particle with spin $J$ at the
ground state so there are $n^3 m^3$ particles.
The center of each object is located at one of the huge cells.
At each huge cell there is one object.

The basis of orthogonal states for this system is:
\be \mid J_{z_{1}}, J_{z_{2}}, ..., J_{z_{n^3 m^3}},
 \vec{r_{1}}, \vec{r_{2}}, ...,\vec{ r_{n^3}}>\ee
where $J_{z_{i}}$ is the spin of the $i$ particle in the $z$ direction so
there are $2J+1$ possibilities for each particle.
$\vec{r_{i}}$ is the location of the center of the $i$'th object at the huge
 cell.
If we move one object a distance equal or bigger then $a$ we obtain a new
orthogonal state.
Thus,
\begin{figure}
\begin{picture}(300,210)(0,0)
\put(50,20){\line(1,0){160}}
\put(210,180){\line(0,-1){160}}
\put(50,180){\line(0,-1){80}}
\put(70,180){\line(0,-1){80}}
\put(90,180){\line(0,-1){80}}
\put(110,180){\line(0,-1){80}}
\put(130,180){\line(0,-1){80}}
\put(50,180){\line(1,0){80}}
\put(50,160){\line(1,0){80}}
\put(50,140){\line(1,0){80}}
\put(50,120){\line(1,0){80}}
\put(50,100){\line(1,0){80}}
\put(150,180){.}
\put(170,180){.}
\put(190,180){.}
\put(50,40){.}
\put(50,60){.}
\put(50,80){.}
\put(220,90){\vector(0,-1){70}}
\put(220,110){\vector(0,1){70}}
\put(60,190){\vector(1,0){10}}
\put(60,190){\vector(-1,0){10}}
\put(55,195){b}
\put(225,95){nb}
\end{picture}
\caption{The system}
\end{figure}
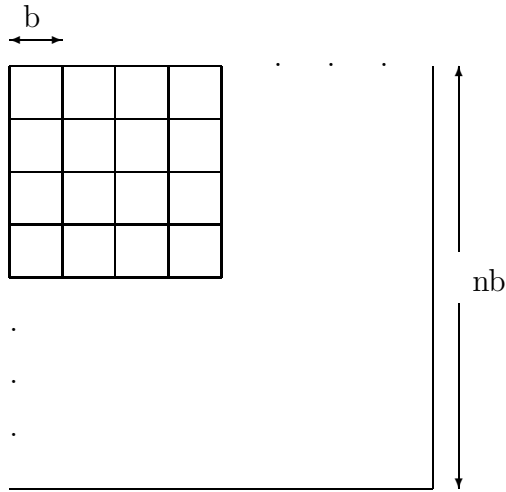
there are $k^3$ possibilities for each huge cell
, where $k=\frac{b}{a}$.
Hence, $[(2J+1)^{m^3}k^3]^{n^3}$ local orthogonal states are required to
describe the system.
Let us examine the validity of the information assumption
 in the absence of gravitation.
Is there a physical apparatus which can distinguish, at least in principle,
 between all the orthogonal states?
The answer to this question is obviously yes (even in the presence of
gravitation) if we use the Von-Neumann measurement
interaction
\be H_{int}=f(t)q\sum \alpha_{i}\mid \phi_{i}> <\phi_{i}\mid,\ee
where the sum is taken over all the orthogonal states  and $q$ is the
canonical position of the apparatus which measures the state of the system.
$\alpha_{i}\neq \alpha_{j}$ for $i\neq j$ so one can find the state of the
system according to the difference in the momentum of the apparatus.
The crucial point is that we can not use the general Von-Neumann interaction
but we must use a local $  H_{int}$.
The correct question is therefore: Is there a physical apparatus which can
distinguish, at least in principle,
between all the orthogonal states by means of a local
interaction?
If one wishes to measure $J_{z_{i}}$ then one should put a detector which
measures the spin of the particle at the $i$'th cell.
In order  to distinguish between all the orthogonal states, one must also
measure the distances between the objects or between an object and the edge
of the huge cell.
In any case the measurement can be carried in the following way:
A clock at one object measures the time $t_{i}$ when a photon is sent
towards the other object.
At the other   object there is a mirror which reflects the photon back to
the first object.
{}From the reading of the clock when the photon arrives, $t_{f}$, one obtains
that the distance between the objects is
\be r=\frac{(t_{i}-t_{f})}{2}\ee
(in units where $c=1$).
Notice that in the absence of gravitation the energy momentum tensor of the
detectors does not
affect the photon thus
there is no difficulty to get
\be \Delta r<a, \;\;\;\;\;\;\forall a\ee
and therefore,  to distinguish between all the orthogonal states .
The information assumption is therefore correct in the absence of gravitation.

{\bf 3.The information loss}

In this section we study the hidden assumption in the definition of information
 in the presence of gravitation.
The special property of gravitation which suggests
that something dramatic might occur in quantum gravity is the
fact that in general relativity, unlike in any other theory,  distances are
 defined by the fields of the theory ($g_{\mu\nu}$).
 Gravitation is, therefore, a geometric force, or in the spirit of
\cite{we} where gravitation is described as a regular force, it has a
universal interaction with all fields  (with the  energy momentum tensor as
the source of interaction).
Thus, the gravitational influence on the particle which is sent in order to
measure the distance between the  objects is universal.
If for instance, one wishes to measure distances in the present of a strong
fluctuation in the electromagnetic field then one should use a chargeless
particle which does not interact with the electromagnetic field,
so the information assumption is correct in Q.E.D.
However, all particles interact with gravitation.
This implies that in quantum gravity the uncertainty in the distance
between the object is larger than the standard uncertainty given by the
Heisenberg relations \cite{Sunny}.

Consider the same physical apparatus which is supposed to
distinguish between all the orthogonal states of the system.
Recall that by the definition of the system $a\gg max[x_{c},x_{d}]$, thus
according to assumption 2 we get (in units where $\hbar=1$)
\be \Delta p_{i}\geq\frac{1}{a}\ee
where $\Delta p_{i}$ is the uncertainty of the momentum of the detector
which measures the spin of the $i$'th particle.
In the weak field approximation:
\be g_{\mu\nu}=\eta_{\mu\nu}+h_{\mu\nu} \;,\;\mid h_{\mu\nu}\mid\ll 1\ee
and
\be P_{\mu}=\int d^{3}xT_{0\mu}\ee
where $T_{\mu\nu}$ is the energy momentum tensor.
Thus \be\Delta \int_{cell} d^{3}xT_{0\mu}\geq\frac{1}{a}\ee
Since $a\gg x_{c}$ we can also use assumption 1 and the well known solution of
the field equations in the weak field approximation, to find that (in
units where $G=1$)

\be g_{\mu\nu}(x,t)=\eta_{\mu\nu}+4\int d^{3}x^{'}\frac{S_{\mu\nu}
(x^{'}, t-\mid x-x^{'}\mid )}{\mid x-x^{'}\mid }\ee
where
 \be S_{\mu\nu}=T_{\mu\nu}-\frac{1}{2}\eta_{\mu\nu}T^{\gamma}_{\gamma}.\ee
There must be no correlation between  different detectors so the
uncertainty in the momentum of one object is:
\be \Delta P\geq\frac{\sqrt{m^3}}{a}.\ee
Using eq.(10) we get,

\begin{equation}\Delta^{2}g_{oi}(x,t)\geq 16\sum_{objects}\frac{m^3}
{a^2 (x-x_{object})^{2}}.\end{equation}
The average distance between  nearest objects is $b$ and thus:
\be \Delta g_{oi}\approx \frac{\sqrt{m^3 n}}{ab}\ee
Now, when one tries to measure the distance between nearest objects one
finds that the velocity of the photon is
\be v=\frac{1}{2}(- g_{oi}+\sqrt(4+ g_{oi}^2))\ee
 Since $ g_{oi}\ll 1$ we get
\be\Delta v\approx  \frac{1}{2}\Delta   g_{oi}
\geqx\frac{\sqrt{m^3 n}}{2ab}\ee
Now, if we write explicitly the speed of light in eq.(4) we obtain
\be\Delta r=\Delta v\frac{t_{f}-t_{i}}{2v^2}=r\frac{\Delta v}{v}\ee
Note that the first equality is due to the fact that the average distance
between nearest objects is $b$ so the correlation
distance of $ \Delta g_{oi}$ is also $b$.
The uncertainty of the
distance between nearest objects $r\approx b$ is therefore
\be \Delta r\geqx\frac{ \sqrt{n m^3}}{2a}.\ee
One can no longer claim that $\Delta r$ is as small as one wishes.
If $\Delta r>a$ then one can not distinguish between all the
$[(2J+1)^{m^3} k^3]^{n^3}$ orthogonal states by means of local measurements
{}.
In fact one can only distinguish between
$[(2J+1)^{m^3}(\frac{k}{l})^{3}]^{n^3}$
orthogonal states, where $l\geqx \frac{\sqrt{n m^3}}{2 a^2}$.
An unavoidable conclusion is that the hidden assumption in the  definition
of information and the general properties of quantum gravity which we assume
can not coexist.
The loss of information is due to the fact that the physical information
that one can obtain about this system (by physical information we mean the
$log$ of the maximal number of orthogonal distinguishable states) is smaller
then the mathematical information of the system (the $log$ of the number of
orthogonal states).
Denoting by $\Delta S$ the information loss, we find:
\be \Delta S\equiv S_{math}-S_{phy}\geqx n^3 log[(2J+1)^{m^3} k^3] -n^3
log[(2J+1)^{m^3} (\frac{k}{l})^3]=3 n^3 log l\ee
Note that $\Delta S$ does not depend on $J$ as Bekenstein-Hawking entropy
is independent of the number of fields.

Another way to illustrate the information loss is to notice that one can
change $a$.
One can start with
$a\geqx (n m^3)^{\frac{1}{4}}$
so  one can distinguish between all the
$[(2J+1)^{m^3} k^3]^{n^3}$
orthogonal states; if the system is changed adiabatically until
$a\leqx (n m^3)^{\frac{1}{4}}$
then one can only distinguish between
$[(2J+1)^{m^3}(\frac{k}{l})^{3}]^{n^3}$
orthogonal states.
If  $a$ is changed back to a region where
$a\geqx (n m^3)^{\frac{1}{4}}$
then there are subsets of states which are distinguishable at the
beginning and at the end but at the middle (when $a\leqx (n m^3)^
{\frac{1}{4}}$)
they are indistinguishable. This suggests that there is no complete
correlation between initial and final distinguishable states so there is no
$S$ matrix and unitarity is violated.
In general, a similar process occurs in black hole evaporation.
At the beginning when there is a collapsing star one can distinguish,
in principle, between almost all the possible initial states of the star
which lead to a black hole.
At the end there is a thermal radiation and again one can distinguish
between all the possible final states.
However in the middle, when there is a black hole one can only distinguish
between subsets which are characterized by the total mass, charge and
angular momentum.

A few remarks are in order:
First there is no local information loss in this system- an
observer who is located at one of the objects will have no difficulty to
obtain all the information about the object.
There is only global information loss in this system.
The part which is lost is the information about the distances between the
objects.
Second, we must make sure that there are no black holes in the system. In
order to do so we should compare the size of the object to its energy and
the size of any subsystem  to its energy.
Since the minimal energy of a particle/detector with uncertainty $a$ is
$\frac{1}{a}$ we find that the total energy of the object is
$E_{ob}\geq\frac{m^3}{a}$ and that the total energy of the system is
$E_{sys}\geq\frac{n^3 m^3}{a}$.
Recall that the size of the object is $ma$ and that the size of the system
is $nb$  so in order to avoid black holes we must impose
\be a>m,\;\;\;\;\;ba>m^3 n^2,\ee
Note that $b$ does not appear in $\Delta S$ so there is no difficulty to
fulfill eq.(20) and avoid black holes.
Third, the loss of information concerning the location of the huge cell $\Delta
r$
is small
comparing to its possible location $b$.
This can be seen easily from the following arguments:
\be \frac{\Delta r}{b}=\frac{\sqrt{n m^3}}{ab}<b^{-\frac{3}{4}},\ee
where we use eq.(20). Since $b\gg 1$ we obtain
\be \frac{\Delta r}{b}\ll 1.\ee
It seems ,therefore that in order to obtain $\Delta r\approx b$ we must
include black holes.
Our goal in this paper is only to present the conceptual problem in quantum
gravity and for this purpose $\Delta r>a$ is enough.

{\bf 4.Conclusions}

For the last few hundred years physics has been described by mathematics.
This description is meaningless unless there is a dictionary which connects
between the mathematical description of physics and physics.
Such a dictionary is absent in quantum gravity if the two assumptions that we
made in the introduction are correct, because although the system which we
study appears to have (according to our assumption)  a well defined
mathematical description (eq.(2)) the physical information contained in the
system is smaller then the mathematical information (eq.(19)).
In our opinion, assumption (2) is incorrect and quantum gravity is a
non-local theory at any scale though there are states with local
interpretation.
In other words, $d$ is not a constant of nature, but rather a function of the
state
of the system.
There are states for which $d$ is smaller than any of the scales in the system
(in general this is the state of the universe now), but as we see there are
states for which $d$ is bigger than some of the scales in the system
although all the scales in the system are much larger then Planck scale.

A natural question which arises now is whether or not this problem is relevant
to the description of  gravitational effects in particle physics.
On the one hand the systems which we consider are so large and complex that one
probably should  not take them into account in particle physics as
virtual states, so the non-locality of those states at large distances
should not affect the large distance locality of particle physics.
On the other hand  the system which we consider is so large and complex
only because we want to stay on  solid ground by using  the weak field
approximation.
When one considers states for which the weak field approximation is not
valid one might obtain information loss in smaller and simpler systems.
In particular, the understanding of this conceptual problem might be crucial
for the description of quantum black holes.
Furthermore, if indeed assumption two is incorrect then the whole
description of quantum gravity is  different than what we are used
to in local \footnote{Local at least at scales much larger than $L_{P}$.}
quantum field theory.
Hence the questions one can ask in quantum gravity
might be different from the questions one asks in local fields theory,
much as the questions in quantum mechanics are not similar to the questions in
classical mechanics.

\vspace{1.5cm}

I am grateful to  Prof. Y.Aharonov Prof. A.Casher and Prof. F.Englert
for helpful discussions.

\end{document}